\documentclass[aps,prd,amsfonts,nofootinbib,supercriptaddress]{revtex4}
\pdfoutput=1

\usepackage{bm}
\usepackage{amsfonts}
\usepackage{subfigure}
\usepackage{feynmp}
\usepackage{amsmath}
\usepackage{amssymb}
\usepackage{graphicx}
\usepackage{hyperref}
\usepackage{color}
\newcommand{\vc}[1]{{\textbf{#1}}}
\newcommand{\mc}[1]{\mathcal{#1}}

\begin{document}

\title{Renormalizing a Viscous Fluid Model for Large Scale Structure Formation}

\author{Florian F\"uhrer$^a$}
\author{Gerasimos Rigopoulos$^b$}

\affiliation{\vspace{0.2cm} $^a$ Institut f\"ur Theoretische Physik, Philosophenweg 16,\\
Universit\"at Heidelberg, 69120 Heidelberg, Germany\\
$^b$ School of Mathematics and Statistics, 
Newcastle University, \\
Newcastle upon Tyne, NE1 7RU, UK}

\begin{abstract}
\noindent

Using the Stochastic Adhesion Model (SAM) as a simple toy model for cosmic structure formation, we study renormalization and the removal of the cutoff dependence from loop integrals in perturbative calculations. SAM shares the same symmetry with the full system of continuity$+$Euler equations and includes a viscosity term and a stochastic noise term, similar to the effective theories recently put forward to model CDM clustering. We show in this context that if the viscosity and noise terms are treated as perturbative corrections to the standard eulerian perturbation theory, they are necessarily non-local in time. To ensure Galilean Invariance higher order vertices related to the viscosity and the noise must then be added and we explicitly show at one-loop that these terms act as counter terms for vertex diagrams. The Ward Identities ensure that the non-local-in-time theory can be renormalized consistently. Another possibility is to include the viscosity in the linear propagator, resulting in exponential damping at high wavenumber. The resulting local-in-time theory is then renormalizable to one loop, requiring less free parameters for its renormalization. 
\end{abstract}

\maketitle

\section{Introduction}
Understanding the evolution of density fluctuations under the influence of gravity is a central, but still open issue in cosmology. 
While fluctuations on large scales are small and can be well described using linear perturbation theory, on small scales the fluctuations grow large and the Standard Perturbation Theory (SPT) fails, see \cite{Bernardeau2001} for a classic review. Two possible sources of the failure of SPT have been identified. The first relates to the influence of very
long-wavelength modes, still within the perturbative regime, on very small scales. After \cite{Crocce2006a,Crocce2006b} first presented their resummation scheme, a lot of progress has been made towards a better understanding of how long-wavelength modes effect smaller scales \cite{Matarrese2007,Valageas2007,Matsubara2008,Bernardeau2013,Anselmi2011,Anselmi2012,Crocce2012},
but also different resummation schemes have been explored
\cite{Pietroni2008,Blas2014}.
The effect of long-wavelength modes on very short modes is absent for equal time correlators due to Galilean Invariance (GI) \cite{Frieman1995,Kehagias2013a,Peloso2013,Blas2013}. Nevertheless, it is still possible that long-wavelength modes can strongly affect intermediate modes and one can hope that the effect of the former on the latter can be inferred from the effect on short modes by adopting a Galilean invariant resummation scheme (see the appendix of \cite{Anselmi2012} for one possible approach).

The second source of failure, that large density fluctuations on small scales
can in principle have a sizable effect on the small fluctuations on
large scales, must be investigated independently of the importance of long-wavelength modes. On small scales not only is the density contrast large but also, and crucially, the single-stream approximation fails and hence the fluid
approximation fails. Calculating the effect of small on large scales is therefore not possible within the framework of SPT or Lagrangian Perturbation Theory (LPT).
To address the issue, it has been proposed to average over small scales such that one is left with equations for the large scale density contrast $\delta$ and velocity $\frac{d\mathbf{x}}{dD}=\mathbf{w}$ plus
an additional effective stress tensor $\mathbf{\sigma}$ which encodes the small
scale dynamics \cite{Buchert:1997dr,Baumann2012,Pietroni2012}. The Euler equation then takes the form 
\begin{align}
\partial_D \mathbf{w}+\frac{3}{2}\frac{\gamma}{D}\left(\mathbf{w}+\nabla
\Phi\right)+\mathbf{w}\cdot\nabla\mathbf{w} &=\frac{\nabla \cdot((1+\delta)
\boldsymbol{\sigma})}{1+\delta}\label{eq:euleraverage},
\end{align}
where $D$ is the linear growth factor, the background cosmology is
encoded in $\gamma=\frac{\Omega_m}{\left(\frac{d\ln(D)}{d
\ln(a)}\right)^2}$ and
$\Phi$ is the rescaled gravitational potential, determined by the Poisson equation $\nabla^2\Phi=\frac{\delta}{D}$. For a $\Lambda$CDM cosmology we have $\gamma \approx 1$. The effective stress $\boldsymbol{\sigma}$ is unknown. One possible strategy of treating this term is to measure it directly from N-body simulations, in which case $\boldsymbol{\sigma}$ acts as a source for the perturbations on large scales \cite{Pietroni2012,Manzotti2014}. Another possible strategy is to attempt its parametrization in terms of the velocity and the density contrast as follows 
\begin{align}\label{EMMT}\frac{\nabla \cdot((1+\delta)
\boldsymbol{\sigma})}{1+\delta}=\mathbf{J}-\frac{\nu_1}{D}\nabla\delta+\nu_2
\nabla^2 \mathbf{w}+\nu_3 \nabla\times\nabla\times \mathbf{w}+\ldots\,, 
\end{align}
where the
$\nu_i$ are effective viscosity parameters and $\mathbf{J}$ is a stochastic
noise. The ellipses denote terms which are higher order in fields
or derivatives which are expected to be suppressed compared to the leading terms. This second strategy goes under the name of Effective Field Theory of Large
Scale Structure (EFToLSS) and has, in its eulerian formulation, drawn a lot of attention in recent years
\cite{Carrasco2012,Mercolli2014,Carrasco2014a,Pajer2013a,Hertzberg2014,Carroll2014,Baldauf2014,Angulo2014}, including the possibility to resum long-wavelength modes \cite{Senatore2015}. A Lagrangian version has also been formulated in
\cite{Porto2014,Vlah2015}. The unknown parameters $\nu_i$ are to be fitted to observations or simulations. 

The added effective terms not only encode the small scale dynamics, they also ensure that physical quantities, for example correlation functions, are independent of the arbitrary smoothing scale. These terms could be treated perturbatively, employing a power counting where the linear propagator does not involve any damping due to the viscosity terms - see for example (\ref{prop-non-local}) for this kind of propagator in the model considered here. The EFToLSS approach effectively corresponds to such a power counting applied ot the full set of fluid equations. As we argue below, to ensure cutoff independence of correlation functions with such a power counting the r.h.s. of (\ref{EMMT}) must be non-local in time. Galilean Invariance of this non-local-in-time theory also dictates the inclusion of higher order non-local-in-time terms of a certain form on the r.h.s. of (\ref{EMMT}). As we will see, these terms ensure cutoff independence and renormalization is possible and consistent with GI. In the EFToLSS literature, the physical meaning of this non-locality in time is traced to the lack of a fast timescale for short wavelengths. It should be noted however that fitting the Power Spectrum obtained by an EFToLSS calculation to a fully non-linear Power Spectrum is possible with a similar accuracy using both local and non-local counter terms \cite{Foreman2015}.

Another possibility is to include the effective viscosity terms in the linear propagator, see (\ref{prop-local}). This power-counting scheme was employed in \cite{Rigopoulos2015a}. The authors of \cite{Blas2015} have used a closely related approach by including a local viscosity and sound speed in the linear fluid perturbation equations, and find that better agreement with N-body simulations can be achieved compared with the usual SPT calculation. We show that in this case the theory is Galillean invariant, local in time and one-loop renormalizable. A  theory local in time is expected if the effective terms are dominated by sufficiently small scales, where, according to the gravitational
free-fall time $\Delta D \sim \delta^{-1/2}$, the typical time scale is much
smaller then the time scale on large scales. This is in particular the case if the
cutoff of the theory is given by the scale where multi-streaming becomes
relevant. For example the viscosity is then given by the microscopic viscosity
plus a contribution from scales beyond the cutoff.

Let us emphasize that the effective terms aim to encapsulate the influence on large scales of
the highly non-linear evolution of short wavelength perturbations, accounting for short wavelength deviations from a single stream fluid. They are not meant to include the effect of a possible non-trivial background phase space distribution or an initial deviation from a single stream fluid, as is the case for free-streaming particles like neutrinos. Nevertheless, the fluid description of such free-streaming particles on scales larger then the free-streaming scale $k_{\rm{FS}}$ can also be interpreted as an Effective Field Theory with $k_{\rm{FS}}$ as the cutoff. Let us briefly sketch this idea. For $n>1$
moments of the velocity distribution are of the order $\overline{w^n}\sim
k_{\rm{FS}}^{-n}$ and thus can be treated as perturbative corrections to
$\delta$ and $\mathbf{w}$. The first corrections are then given by the stress
tensor, which obeys, neglecting the third moment,
\begin{align}
\partial_D
\boldsymbol{\sigma}+3\frac{\gamma}{D}\boldsymbol{\sigma}+\mathbf{w}\nabla\cdot\boldsymbol{\sigma}+\boldsymbol{\sigma}\cdot\nabla\mathbf{w}+(\nabla\mathbf{w})^{T}\cdot\boldsymbol{\sigma}=0.
\end{align}
Splitting the stress into a background and a perturbation
$\mathbf{\sigma}=\bar{\sigma}\mathbf{1}+\delta\boldsymbol{\sigma}$ one can write the
stress, for $\gamma=1$, as
\begin{align}
\boldsymbol{\sigma}=\mathbf{1}\bar{\sigma}_i
\left(\frac{D_i}{D}\right)^3+\delta\boldsymbol{\sigma}_i\left(\frac{D_i}{D}\right)^3+\mathbf{1}\bar{\sigma}_i
\left(\frac{D_i}{D}\right)^3\int_{D_i}^D d\eta\:(\nabla \mathbf{w}+(\nabla
\mathbf{w})^{T})(\eta)+\ldots,
\end{align}
with $\bar{\sigma}_i$ and $\delta\boldsymbol{\sigma}_i$ being the
background value and the perturbation of the stress at some initial time $D_i$.
Plugging this into the right hand side of the Euler equation \ref{eq:euleraverage}
we have
\begin{align}
\frac{\nabla \cdot((1+\delta)
\boldsymbol{\sigma})}{1+\delta}=\nabla\cdot\delta\boldsymbol{\sigma}_i\left(\frac{D_i}{D}\right)^3+\bar{\sigma}_i\left(\frac{D_i}{D}\right)^3\nabla\delta
+\mathbf{1}\bar{\sigma}_i
\left(\frac{D_i}{D}\right)^3\int_{D_i}^D d\eta\:(\nabla^2
\mathbf{w}+\nabla\nabla\cdot \mathbf{w})(\eta)+\ldots.
\end{align}
To this order the initial stress perturbation plays the role of a
stochastic noise and the background stress induces a local sound speed and
a non-local-in-time viscosity. Higher order contributions can be obtained
straightforwardly in the double expansion in  $\frac{\nabla}{k_{\rm{FS}}}$ and
the fields $\delta$ and $\mathbf{w}$. Note that the time dependence of the
effective terms is fixed and does not coincide with the SPT loop time dependence so
the theory cannot be renormalized. Since the velocity moments of the background
and the initial distribution are in principle known, they can be resumed such
that one obtains a theory which is non-local and valid at
all scales as long as the density contrast is small
\cite{Fuhrer:2014zka}. We see that the effective long wavelength theory for CDM may be thought off as analogous to this approach to free-streaming particles. The analogy is imperfect though, given that different time dependences of the effective terms are required for CDM.     

The outline of our paper is as follows. In section \ref{sec:StAdMo} we introduce our simplified toy model, the Stochastic Adhesion Model. Then in section \ref{sec:Gal} we discuss GI and how it constrains the allowed terms that can be used to parameterize the effective stress tensor. In section \ref{sec:renom}
we discuss renormalization of the local-in-time and non-local-in-time versions of the effective theory. We conclude in section \ref{sec:conclusion}.


\section{The stochastic adhesion model}\label{sec:StAdMo}
Instead of discussing the full set of equations consisting of the
continuity and the Euler equation, we study the technically
simpler Stochastic Adhesion Model (SAM) as a toy model. The SAM, as already discussed in \cite{Rigopoulos2015a}, can be obtained from the fluid equations by a Zeld'ovich
approximation, see also \cite{Gurbatov1989,Buchert1999,Matarrese2002,Gaite2012}
for earlier work on the Adhesion Model and the Burgers Equation. The Zel'dovich approximation reads $\mathbf{w}=-\nabla\Phi$ and as a result $d\mathbf{w}/dD = \partial_D \mathbf{w} + \mathbf{w}\cdot\nabla\mathbf{w}=0$. This approximation decouples the Euler equation from the continuity equation. In principle one can obtain $\delta$ from the continuity equation once $h$ is known, for details see \cite{Rigopoulos2015a}. In the following we will not consider the continuity equation, since the scope of this paper is to discuss the interplay between renormalization and GI, which can be done by considering the Euler equation alone.  The SAM is obtained by writing $d\mathbf{w}/dD =\partial_D \mathbf{w} + \mathbf{w}\cdot\nabla\mathbf{w}=\frac{\nabla \cdot((1+\delta) \boldsymbol{\sigma})}{1+\delta}$ and expressing the effective stress tensor as discussed above. It thus approximates deviations of fluid elements from their long wavelength Zel'dovich trajectories. One ends up with a time dependent Kadar-Parisi-Zhang (KPZ) equation \cite{Kardar1986} for the velocity potential $h$, defined by $\mathbf{w}=-\nabla h$.

SAM is a stochastic field theory and is most conveniently formulated by defining the MSRJD action\footnote{Named after Martin, Siggia, Rose, Jansen and De Dominicis \cite{Altland:2010si}.} 
\begin{align}\label{eq:action}
S &=\frac{1}{2}\int dDdD'\frac{d^3k}{(2\pi)^3}\,\left[ \left(\begin{smallmatrix}h_{\vc{k}}\,, &\chi_{\vc{k}} \end{smallmatrix}\right)_D\left(\begin{smallmatrix}0 & \left(-\partial_D+\nu k^2\right)\delta(D-D')\\ \left(\partial_D +\nu k^2\right)\delta(D-D') & {\rm i}\mc{N}(k,D,D')\end{smallmatrix}\right)
\left(\begin{smallmatrix}h_{-\vc{k}} \\ \chi_{-\vc{k}} \end{smallmatrix}\right)_{D'} + \mc{L}_{\rm int}\right]\\\nonumber
 &= \frac{1}{2}\int_{DD'}\frac{d^3k}{(2\pi)^3}\,\left[ \left(\begin{smallmatrix}h_{\vc{k}}\,, &\chi_{\vc{k}} \end{smallmatrix}\right)\hat{G}_0^{-1}
 \left(\begin{smallmatrix}h_{-\vc{k}} \\ \chi_{-\vc{k}} \end{smallmatrix}\right) + \mc{L}_{\rm int}\right]
\,,
\end{align}
where 
\begin{align}
\hat{G}_0^{-1}=\left(\begin{smallmatrix}0 & \left[G^A_0\right]^{-1}\\ \left[G^R_0\right]^{-1} & {\rm i}\mc{N}(k)\end{smallmatrix}\right)\,.
\end{align}
and
\begin{align}\label{intraction}
\mc{L}_{\rm int}=\int \frac{d^3q_1}{(2\pi)^3}\frac{d^3q_2}{(2\pi)^3}\,\,\left(\vc{q}_1\cdot\vc{q}_2\right)\,\,\chi_{\vc{k}}h_{\vc{q}_1}h_{\vc{q}_2}\,\delta(\vc{k}+\vc{q}_1+\vc{q}_2)
\end{align}
is the interaction vertex. In the second line of (\ref{eq:action}) we condensed the notation for time integrations. The field $\chi$ is an auxiliary field and the $\chi^2$ term in the action encodes stochasticity. It can be used to encode both stochastic initial conditions as well as the continuously acting stochastic part of the tress tensor, modeling the action of small scale fluctuations.  Accordingly, $\mathcal{N}$ contains the initial Power Spectrum
$P_{\Phi_{\rm in}}$ as well as the Power Spectrum $\Delta=\langle J J\rangle$ of the
(gaussian) noise $J$.
For the Fourier transform of $\mathcal{N}$ we assume
\begin{align}
\mathcal{N}(D,D';\mathbf{k},\mathbf{k}')=P_{\Phi_{\rm{in}}}(k)(2\pi)^3\delta_D(\mathbf{k}+\mathbf{k}')
\delta_D(D-D_{\rm{in}})\delta_D(D'-D_{\rm{in}})+\Delta(D,D')(2\pi)^3\delta_D(\mathbf{k}+\mathbf{k}'),
\end{align}
with the scale independence of $\Delta$ ensuring that the small scale fluctuations
induce the well known $k^2$ peculiar velocity Power
Spectrum at large scales and, correspondingly the $k^4$ tail in the density. The term $\nu\nabla^2h$ is the effective viscosity term. $G^{R(A)}_0$ is the \emph{free} Retarded (Advanced) Green function and the notation $\left[G^{R(A)}_0\right]^{-1}$ is used to denote the operators appearing in (\ref{eq:action}) with Retarded (Advanced) boundary conditions. All correlators of interest can then be obtained from the generating functional 
\begin{align}
\mathcal{Z}=\int Dh D\chi \,\,e^{iS}\,,
\end{align}
with the MSRJD propagator $\hat{G}$ defined as the functional and matrix inverse of the matrix in the quadratic part of the MSRJD action: 
\begin{align}
\hat{G}_0=
\left(\begin{smallmatrix}
F_0(D,D';k)&-{\rm i}G_0^R(D,D';k)\\
-{\rm i}G^A_0(D,D';k)&0\end{smallmatrix}\right)=
\left(\begin{smallmatrix}\langle h_{\vc{k}}(D)h^\star_{\vc{k}}(D')\rangle & \langle h_{\vc{k}}(D)\chi^\star_{\vc{k}}(D')\rangle  \\ \langle \chi_{\vc{k}}(D)h^{\star}_{\vc{k}}(D')\rangle & \langle \chi_{\vc{k}}(D)\chi^\star_{\vc{k}}(D')\rangle \end{smallmatrix}\right).
\end{align} 
The advanced and retarded Green functions are not independent since $G_0^R(D,D') = G^{A\star}_0(D',D)$. Furthermore, knowledge of $G_0^R$ allows the computation of $F_0$ as
\begin{align}
F_0(D,D';k)=\int\limits_0^D dudv \,\,G_0^R(D,u;k)\mc{N}(u,v;k) G_0^A(v,D';k).
\end{align}  
 
\begin{figure}
\centering
\includegraphics[width=0.99\textwidth]{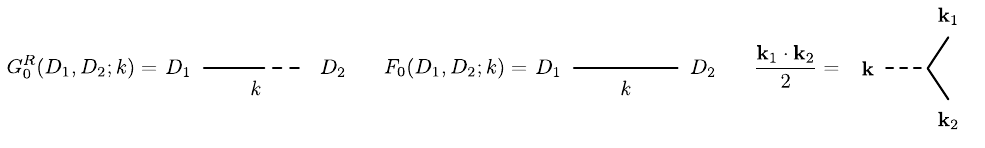}
\caption{The Feynman rules for the action \eqref{eq:action} - for more details see \cite{Rigopoulos2015a}. As usual, wave-vector
conservation applies and the vertex has to
be integrated over internal momenta and time.}
\label{fig:rules}
\end{figure}

As we already discussed, the theory is expected to be non-local in time and this can be implemented by the replacement $\nu(D)\nabla^2h(D)\rightarrow \int_{D_{\rm{in}}}^D dD' \: \nu(D,D') \nabla^2 h(D')$.
From the action \eqref{eq:action} one can read off the Feynman rules depicted in
figure \ref{fig:rules}. Note, that the form of the propagators
$G_0^R(D,D';\mathbf{k})$ and $F_0(D,D';\mathbf{k})$ is different if the viscosity and the noise are treated pertubatively (\'a la EFToLSS) or non-perturbatively. Note also that $F_0$ is simply the linear Power Spectrum $P_{\rm L}$ of $h$.

The MSRJD action and propagator imply that the "self-energy" (the sum of all 1PI diagrams\footnote{1PI-diagrams are those
	diagrams which cannot be cut into two by cutting a single line.}) also has the structure  
\begin{align}
\hat{\Sigma} =\left(\begin{array}{cc}
0 & \Sigma^A(D,D';k) \\ 
\Sigma^R(D,D';k) & i \Phi(D,D';k)
\end{array}\right)\,.
\end{align}   
If $G$ is the full \emph{dressed} Green function it satisfies the Schwinger-Dyson equation 
\begin{align}
\left(\hat{G}_0^{-1} -\hat{\Sigma}\right)\circ\hat{G}=\hat{1}
\end{align}
where the circle product denotes integrations over time and matrix multiplications. Writing it explicitly we obtain
\begin{align}
\left(\partial_D +\nu k^2 \right) G^R(D,D') - \int du \, \Sigma^R(D,u) G^R(u,D') = \delta\left(D-D'\right)\,, \label{SD1}\\
\left(-\partial_D +\nu k^2 \right) G^A(D,D') - \int du \, \Sigma^A(D,u) G^A(u,D') = \delta\left(D-D'\right)\label{SD2}\,,\\
\left(\partial_D +\nu k^2 \right) F(D,D') - \int du \, \Sigma^R(D,u) F(u,D') + \int du \left( \mc{N}(D,u) - \Phi(D,u)\right)G^A(u,D')= 0\label{SD3} \,.
\end{align}    

To close this section, we emphasize that SAM is not intended as a tool for precision cosmology calculations as it contains uncontroled approximations. Nontheless, it can reproduce qualitatively 
the morphological structure of the cosmic web obtained from N-Body simulations - see \cite{Weinberg1990,Kofmann1992,Nusser1990} for some early works on the adhesion model. The addition of a stochastic component could be used to parametrize short scale, highly non-linear processes and could improve the results of those early works by generating more realistic short scale power. Furthermore, it seems that an irreducible stochastic component is necessary for describing the effects of short scales and becomes dominant over further additions to the stress energy tensor \cite{Baldauf2015}. With these remarks we postpone a detailed evaluation of SAM for future work. What is important for us here is simply that 
the SAM is invariant under Extended Galilean Transformations (GT), like the complete fluid equations, and we are therefore able to discuss the interplay between Galilean Invariance (GI), non-locality in time and renormalization with the SAM as a conceptually useful toy model.


\section{Galilean Invariance}
\label{sec:Gal}
Symmetries constrain the allowed terms
to be added to the fluid equations for CDM. 
In the previous section we already stressed that the fluid equations as well as the SAM are invariant under GT. The GI of the fluid equations in the context of LSS and the corresponding consistency relations were already discussed in \cite{Kehagias2013a,Peloso2013,Peloso2014,Ben-Dayan:2014hsa}. GI of the fluid equations is the symmetry of the relativistic equations with a non-trivial Newtonian limit \cite{Horn2014a} and are therefore related to diffeomorphism invariance of the full relativistic theory. A GT is a time dependent boost with a velocity $\boldsymbol{\beta}(D)$. The coordinates then transform according to
\begin{align}
&D\rightarrow D'=D\\\notag
&\mathbf{x}\rightarrow\mathbf{x}'=\mathbf{x}+\int_{D_i}^D d\eta\: \boldsymbol{\beta}\left(\eta\right)\equiv\mathbf{x}+\mathbf{T}\left(D\right),
\end{align}
and the velocity potential transforms accordingly as
\begin{align}
h(D,\mathbf{x})\rightarrow h(D',\mathbf{x}')=h(D,\mathbf{x}+\mathbf{T}(D))-\mathbf{x}\cdot\boldsymbol{\beta}(D).
\end{align}
The action \eqref{eq:action} then transform as follows
\begin{align}
S[h,\chi]\rightarrow S[h,\chi]+\delta S_{\mathcal{N}}[h,\chi]+\int dD d^3x\:
\chi\left(-\mathbf{x} \cdot \partial_D
\boldsymbol{\beta}+\frac{1}{2}\boldsymbol{\beta}^2\right),\label{eq:actionGT}
\end{align} 
where we used that time derivatives are not GI, but the convective derivative $\partial_{D}+\frac{1}{2}\nabla h\cdot\nabla$ is GI.
Compared to the action in equation \eqref{eq:action}, equation
\eqref{eq:actionGT} contains two extra terms. The term $\delta
S_{\mathcal{N}}[h,\chi]$ will vanish for a GI stochastic noise. The term $\int dD
d^3x\: \chi\left(-\mathbf{x} \cdot \partial_D
\boldsymbol{\beta}+\frac{1}{2}\boldsymbol{\beta}^2\right)$ contains two unobservable
contributions. The $\boldsymbol{\beta}^2$ terms simply adds a constant contribution to the
velocity potential, while the $\mathbf{x}\cdot\partial_D \boldsymbol{\beta}$ term is a frame
fixing term and ensures that the homogeneous mode of the velocity is given by
$\boldsymbol{\beta}$ \cite{Peloso2013}.

As discussed above,  higher order terms are allowed but GI only allows terms built of second or higher derivatives of $h$.
For example we can add a second order viscosity
\begin{align}
\nu^{(2)} \nabla^4 h
\end{align} 
or new vertices as
\begin{align}
\lambda \left(\nabla^2 h\right)^2+g \left(\partial_i \partial_j h\right)\left(\partial^i \partial^j h\right).\label{eq:vertexsecondorder}
\end{align}
Let us now have look at the noise and how it is constrained by GI. The Power
Spectrum of a statistically homogeneous and isotropic noise $J$ is of the form $\langle
J(D_1,\mathbf{x}_1) J(D_1,\mathbf{x}_2)\rangle=\Delta(D_1,D_2,|\mathbf{x}_1-\mathbf{x}_2|)$. The noise term in the action is only GI if the noise Power Spectrum is invariant
\begin{align}
\Delta(D_1,D_2,|\mathbf{x}_1-\mathbf{x}_2|)=\Delta(D_1,D_2,|\mathbf{x}_1+\mathbf{T}(D_1)-\mathbf{x}_2-\mathbf{T}(D_2)|),
\end{align}
which is only the case if the noise is temporally white
\begin{align}
\Delta(D_1,D_2,|\mathbf{x}_1-\mathbf{x}_2|)=\Delta(D_1,|\mathbf{x}_1-\mathbf{x}_2|)\delta_D(D_1-D_2).
\end{align}
Similar arguments hold for higher order correlators or a multiplicative noise.

A noise with a finite correlation time is not apparently GI.
However, consider the noise term evaluated along the path
of a fluid element \cite{Carrasco2014a}
\begin{align}
J\left(D;\mathbf{x}_{\rm{fl}}\left(D,D_i\right)\right)\label{eq:nonlocalnoise},
\end{align}
where the position of the fluid element can be obtained by solving
\begin{align}
\mathbf{x}_{\rm{fl}}\left(D,D'\right)=\mathbf{x}+\int_{D'}^D d\eta\: \nabla h \left(\eta,\mathbf{x}_{\rm{fl}}\left(D,\eta\right)\right).
\end{align}
Since $\mathbf{x}_{\rm{fl}}$ does not change under GTs, neither does the argument of $J$ and its correlators are invariant. It is interesting to note that a noise of the form given in equation \eqref{eq:nonlocalnoise} is nothing but a solution of an equation of the form:
\begin{align}\label{eq:noise}
\frac{d J}{dD}\Big|_{\mathbf{x}_{\rm fl}}=\partial_{D} J+\mathbf{w}\cdot\nabla J=\ldots,
\end{align} 
i.e. of a derivative taken along the fluid flow lines. The ellipses denote
possible further terms consistent with Galilean symmetry. So a non-local noise
can be seen as new degree of freedom, governed by \eqref{eq:noise}, which can be added to the set of
equations. As we will discuss
in section \ref{sec:renom} the time dependence of the noise is fixed by the
time dependence of the loops. This means that it is sufficient to provide initial
conditions to specify the noise, suggesting that a non-local noise arises from
coarse graining over the initial conditions.\footnote{Our discussion of free-streaming particles in the
introduction provides another example where the initial stress tensor has
been integrated out.}

If we treat the dependence on the fluid path perturbatively we find for the noise
\begin{align}
J(D;\mathbf{x}_{\rm{fl}}(D,D_i))&=J(D;\mathbf{x})+\int^D_{D_i}d\eta\:\nabla h
(\eta;\mathbf{x})\cdot\nabla J(D;\mathbf{x})+\ldots.
\end{align}
For the  corresponding Power Spectrum we find a temporal non-white noise plus corrections in form of a multiplicative noise
\begin{align}\label{eq:non-loc-noise}
&\Delta\left(D_1,D_2;\left|\mathbf{x}_{\rm{fl}}(D_1,D_i)
-\mathbf{x}_{\rm{fl}}(D_2,D_i)\right|\right)=\Delta\left(D_1,D_2;\left|\mathbf{x}_1-\mathbf{x}_2\right|\right)\\
+&\int^{D_1}_{D_i}d\eta\:(\nabla
h)(\eta;\mathbf{x}_1)\cdot\nabla_1\Delta\left(D_1,D_2;\left|\mathbf{x}_1-\mathbf{x}_2\right|\right)+\int^{D_2}_{D_i}d\eta\:(\nabla
h)(\eta;\mathbf{x}_2)\cdot\nabla_2\Delta\left(D_1,D_2;\left|\mathbf{x}_1-\mathbf{x}_2\right|\right)+\ldots\notag.
\end{align}
So a non-local noise is allowed as long as it is the first term in a series of terms all with the \emph{same} coefficient function.\\
Evaluating the fields along the fluid path $\mathbf{x}_{\rm fl}$ we can likewise
generalize local terms containing $h$ to non-local terms. For example a
non-local viscosity reads
\begin{align}\label{eq:non-loc-viscosity}
\int_{D_{\rm{in}}} ^D dD'\: \nu(D,D') \nabla^2
h(D';\mathbf{x}_{\rm{fl}}\left(D',D_i\right))&=\int_{D_{\rm{in}}} ^D dD'\:
\nu(D,D') \nabla^2 h(D';\mathbf{x})\notag\\
&+\int_{D_{\rm{in}}} ^D dD'\int_{D_{\rm{in}}} ^{D'} d\eta\: \nu(D,D') \nabla^2
\left((\nabla h(\eta;\mathbf{x}))\cdot\nabla h(D';\mathbf{x})\right)\ldots.
\end{align}
Similarly to the noise, a non-local viscosity is allowed as the first term in a series of terms, all with the \emph{same} coefficient.

Observe that these extra terms can and must contain the
velocity itself, so terms with only one derivative acting on $h$ appear in the action. The non-local-in-time terms in \eqref{eq:non-loc-noise} and \eqref{eq:non-loc-viscosity}, a consequence of GI, lead at one loop to the new vertices
\begin{align}
\ \notag\\
\parbox{20mm}{\begin{fmffile}{diagrams/viscosityvertex}
\begin{fmfgraph*}(60,40)
\fmfleft{i}
\fmftop{o1}
\fmfbottom{o2}
\fmf{dashes}{i,v}
\fmf{plain}{v,o1}
\fmf{plain}{v,o2}
\fmflabel{$D,\mathbf{k}$}{i}
\fmflabel{$D_1,\mathbf{k}_1$}{o1}
\fmflabel{$D_2,\mathbf{k}_2$}{o2}
\fmfdot{v}
\end{fmfgraph*}
\end{fmffile}}&=-\frac{k^2\mathbf{k}_1\cdot\mathbf{k}_2}{2}\left(\nu(D,D_1)\theta(D_1-D_2)+\nu(D,D_2)\theta(D_2-D_1)\right)\\
\ \notag\\
\ \notag\\
\ \notag\\
\parbox{20mm}{\begin{fmffile}{diagrams/noisevertex}
\begin{fmfgraph*}(60,40)
\fmfleft{i}
\fmftop{o1}
\fmfbottom{o2}
\fmf{plain}{i,v}
\fmf{dashes}{v,o1}
\fmf{dashes}{v,o2}
\fmflabel{$D,\mathbf{k}$}{i}
\fmflabel{$D_1,\mathbf{k}_1$}{o1}
\fmflabel{$D_2,\mathbf{k}_2$}{o2}
\fmfdot{v}
\end{fmfgraph*}
\end{fmffile}}&=\left(\mathbf{k}\cdot\mathbf{k}_2\Delta(D_1,D_2;k_2)+\mathbf{k}\cdot\mathbf{k}_1\Delta(D_1,D_2;k_1)\right)(\theta(D_1-D)+\theta(D_2-D)),\notag\\
\end{align}
where as usual ``momentum" conservation is implied. In contrast to the usual
vertex one has to integrate over all three times $D$,$D_1$ and $D_2$.
In the EFToLSS approach these terms will only appear at second order, so are relevant for the one-loop bispectrum and the two-loop Power Spectrum.


\section{Renormalization}\label{sec:renom}
We now discuss the implications of GI for the renormalization of UV-divergences.
By a UV-divergence we refer to the leading contribution from hard loop momenta, irrespectively of whether the loop integrals are finite or infinite. In both cases these are unphysical contributions and must be removed from physical quantities by counter terms corresponding to the effective terms. Since in SPT loop integrals are finite for realistic initial conditions, actual divergences 
are not present at low orders, but will arise at higher orders from loops containing the effective terms.

When renormalizing the loop integrals one has to pay attention to the fact that they have a non-trivial time dependence, so the counter terms must match the time dependence of the UV-divergences, either local or non-local in time. Since the time dependence of the effective terms is \emph{not} constrained by GI or any other symmetry, the time dependence can always be chosen as required for the cancellation of UV-divergences, but if the effective terms are non-local in time the same divergence appears at higher order again. Since the counter terms appear again at higher order, these divergences are automatically renormalized. We will explicitly show that this happens at one loop. Afterwards, we will discuss how the Ward-Identities (WI) from GI ensure that this will happen at any loop order.


\subsection{Non local in time}
The approach taken in the EFToLSS leterature treats the effective viscosity and the noise as perturbative corrections. The diagrammatic expansion in the EFToLSS therefore consists of the SPT diagrams plus counter-term diagrams for the effective terms. In the framework of SAM this translates into the free propagators reading
\begin{align}
G^R_0(D,D';k)&=\theta(D-D')\notag\\ \label{prop-non-local}
F_0(D,D';k)&=P_{\Phi_{\rm in}}(k),
\end{align}
where we used initial conditions in the past at $D_{\rm in}\rightarrow
0$. At one loop the self-energies are 
\begin{align}
\Sigma^R\left(D_1,D_2;k\right)&=\quad\quad\quad
\parbox{45mm}{\begin{fmffile}{diagrams/viscosity}
\begin{fmfgraph*}(80,30)
\fmfleft{i}
\fmfright{o}
\fmf{dashes,tension=3}{i,v1}
\fmf{plain,tension=3}{v2,o}
\fmf{plain}{v1,v3}
\fmf{dashes}{v3,v2}
\fmf{plain,left,tension=1}{v1,v2}
\fmflabel{$D,\mathbf{k}$}{i}
\fmflabel{$D',-\mathbf{k}$}{o}
\end{fmfgraph*}
\end{fmffile}}= -\frac{k^2}{12} \theta(D_1-D_2) I_2\notag\\
\Phi\left(D_1,D_2;k\right)=\Phi(k)&=\quad\quad\quad\parbox{45mm}{\begin{fmffile}{diagrams/noise}
\begin{fmfgraph*}(80,30)
\fmfleft{i}
\fmfright{o}
\fmf{dashes,tension=3}{i,v1}
\fmf{dashes,tension=3}{v2,o}
\fmf{plain,right,tension=1}{v1,v2}
\fmf{plain,left,tension=1}{v1,v2}
\fmflabel{$D_1,\mathbf{k}$}{i}
\fmflabel{$D_2,-\mathbf{k}$}{o}
\end{fmfgraph*}
\end{fmffile}}=\frac{1}{4}Y_4(k)-\frac{k^2}{8} Y_2(k)+\frac{k^4}{64}
Y_{0}(k),\label{eq:selfenergiesEFT}
\end{align}
where we defined
\begin{align}
I_n&=\int \frac{d^3q}{(2\pi)^3}\: q^n P_{\Phi_{\rm in}}(q)\\
Y_n(k)&=\int \frac{d^3q}{(2\pi)^3}\: q^n P_{\Phi_{\rm in}}\left(\left|\mathbf{q}-\frac{\mathbf{k}}{2}\right|\right)P_{\Phi_{\rm in}}\left(\left|\mathbf{q}+\frac{\mathbf{k}}{2}\right|\right).
\end{align}
As a consequence of causality, inherent in the MSRJD form of the action, $\Sigma$ is only non-vanishing for $D_2<D_1$, while $\Phi$ is always non-vanishing. Since the self-energies are not sharply peaked for any $D_1$ and $D_2$ nor for $D_1\approx D_2$, they require non-local counter terms.
It is immediately clear
that the dependence of external wave vector $k$ of the leading UV-divergences
of $\Sigma$ matches the $k^2$ dependence of the viscosity. The leading
UV-divergence in $\Phi$ comes from $Y_4$, the term with the highest power of $q$ under the integral.
Taking the limit where the loop momentum $q$ is much larger then the external momentum $k$, $q\gg k$, we see that the divergence can be absorbed  into the scale-independent noise. At higher order it is ensured by the form of the SPT kernels, as dictated by
momentum conversation, that the $k$-dependence of the divergences always
matches the one of the viscosity and the noise \cite{Bernardeau2001,
	Pajer2013a}.

Using the above one-loop self-energies the Schwinger-Dyson equations give
 \begin{align}
 &\partial_D G^R(D,D') + k^2 \int\limits_0^D du \left(\nu\left(D,u\right)+\frac{I_2}{12}\right)G^R\left(u,D'\right) = \delta\left(D-D'\right)\,,\\
 &\partial_D F(D,D') + k^2 \int\limits_0^D du \left(\nu\left(D,u\right)+\frac{I_2}{12}\right)F(u,D')  +  \int\limits_{0}^{D} du \,\left( \mc{N}(D,u, \vc{k}) - \Phi(\vc{k})\right)G^A(u,D')= 0\,.
 \end{align}	
They can be renormalized by constant contributions to the viscosity and noise   
\begin{align}
\nu(D_1,D_2)=c_v^2\notag\\
\Delta(D_1,D_2)=\Delta.
\end{align}
It has been noticed \cite{Carroll2014,Baldauf2014} that at lowest order a non-local viscosity can be mimicked by a local one. In principle this is still possible at higher order but then new terms are necessary correcting for the error made by using a local viscosity instead of a non-local one. For the noise a similar procedure is not possible.
\begin{figure}
\centering
\includegraphics[width=0.99\textwidth]{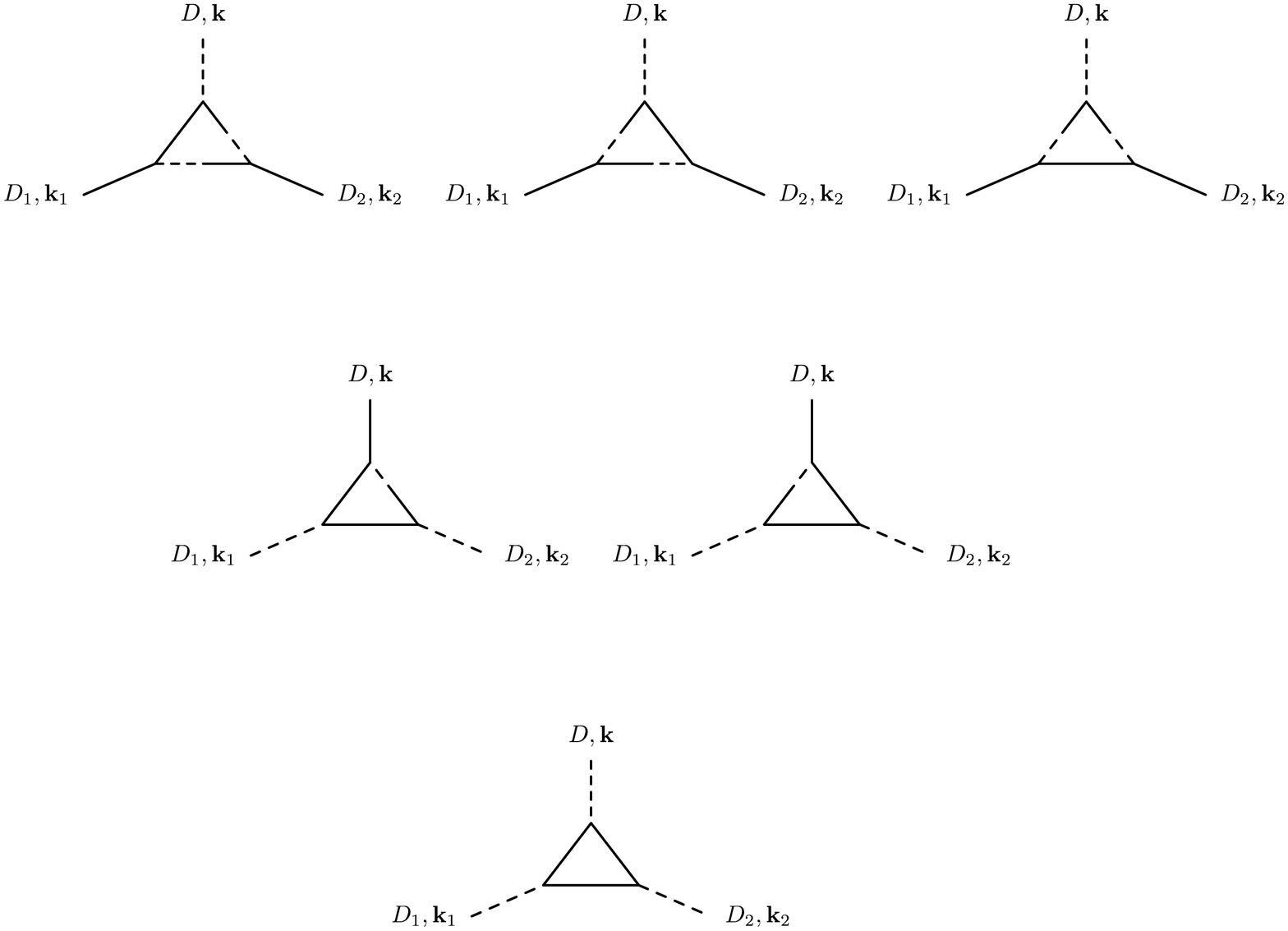}
\caption{The 6 one-loop contribution to the bispectrum. The upper
three renormalize the vertex. The two diagrams in the second line generate a multiplicative noise. The diagram in the last line generates a bispectrum for the noise.}
\label{fig:vertexdiagrams}
\end{figure}

As discussed in section \ref{sec:Gal} the same constants $c_v^2$ and $\Delta$
appear as coefficients of vertices, so the same divergences as in the vertices
must be present in the vertex corrections, which are given by the sum of the three upper diagrams shown in figure \ref{fig:vertexdiagrams}
\begin{align}
\Pi(D,D_1,D_2;\mathbf{k}_1,\mathbf{k}_2)&=-\theta(D-D_1)\theta(D_1-D_2)
\frac{k^2
\mathbf{k}_1\cdot\mathbf{k}_2}{12}I_2+\theta(D-D_1)\theta(D-D_2)\frac{(\mathbf{k}_1\cdot\mathbf{k}_2)^2}{12}I_2+(D_1\leftrightarrow
D_2).\label{eq:Pieft}
\end{align}
Note that contributions to $\Pi$ which would renormalize the vertex
$\frac{\mathbf{k}_1\cdot\mathbf{k}_2}{2}$ cancel among the three different
diagrams.
The first term in equation \eqref{eq:Pieft} has exactly the from required by
GI, the second term can be renormalized by  a non-local version of the $g$-term in equation
\eqref{eq:vertexsecondorder} while the non appearance of a term corresponding to $\lambda$ in \eqref{eq:vertexsecondorder} is a consequence of the simple form of the vertex.

The second class of diagrams we have to consider are those generating a
multiplicative noise (the two diagrams in the second line of figure
\ref{fig:vertexdiagrams}). Their sum is
\begin{align}
\Psi(D,D_1,D_2;\mathbf{k}_1,\mathbf{k}_2)&=\theta(D_1-D)\bigg(-\frac{\mathbf{k}_1\cdot\mathbf{k}}{16}Y_4(k_1)-\frac{\mathbf{k}_2\cdot\mathbf{k}}{16}Y_4(k_2)+\frac{1}{8}\mathbf{k}\cdot\left(\mathbf{Y}_4(k_1)+\mathbf{Y}_4(k_2)\right)\cdot\mathbf{k}\notag\\
&+\frac{k_1^2\mathbf{k_1}\cdot\mathbf{k}-(\mathbf{k_1}\cdot\mathbf{k})^2}{32}Y_2(k_1)+\frac{k_2^2\mathbf{k}_2\cdot\mathbf{k}-(\mathbf{k}_2\cdot\mathbf{k})^2}{32}Y_2(k_2)-\frac{1}{32}\mathbf{k}\cdot\left(k_1^2\mathbf{Y}_2(k_1)+k_2^2\mathbf{Y}_2(k_2)\right)\cdot\mathbf{k}\notag\\
&-\frac{k_1^4\mathbf{k}\cdot\mathbf{k}_1-k_1^2(\mathbf{k}\cdot\mathbf{k}_1)^2}{128}Y_0(k_1)-\frac{k_2^4\mathbf{k}\cdot\mathbf{k}_2-k_2^2(\mathbf{k}\cdot\mathbf{k}_2)^2}{128}Y_0(k_2)\bigg)+(D_1\leftrightarrow
D_2),
\end{align}
where we used that $\int d^3 q\: \mathbf{q} q^n P_{\Phi_{\rm
in}}\left(\left|\mathbf{q}-\frac{\mathbf{k}}{2}\right|\right)P_{\Phi_{\rm in}}\left(\left|\mathbf{q}+\frac{\mathbf{k}}{2}\right|\right)=0$ and defined
\begin{align}
\mathbf{Y}_{n+2}(k)&=\int \frac{d^3q}{(2\pi)^3}\: \mathbf{q}\mathbf{q} q^n
P_{\Phi_{\rm in}}\left(\left|\mathbf{q}-\frac{\mathbf{k}}{2}\right|\right)P_{\Phi_{\rm in}}\left(\left|\mathbf{q}+\frac{\mathbf{k}}{2}\right|\right).
\end{align}
The leading divergences are again those of $Y_4$ and $\mathbf{Y}_4$, where due
to rotational invariance $\mathbf{Y}$ must be of the form
\begin{align}
\mathbf{Y}_{n+2}(\mathbf{k})=F(k)\mathbf{1}+G(k) \mathbf{k}\mathbf{k}.
\end{align}
Since in the limit $q \gg k $, $\mathbf{Y}_{n+2}(\mathbf{k})$ depends only on
the absolute value $k$, we conclude that the leading divergence stems form
$F(k)$, which is the same as that of $Y_4$. So we can simply replace F by
$\frac{Y_{n+2}}{3}$. In that limit $\Psi$ reads
\begin{align}
\Psi(D,D_1,D_2;\mathbf{k}_1,\mathbf{k}_2)&\sim\theta(D_1-D)\left(-\frac{\mathbf{k}_1\cdot\mathbf{k}}{16}Y_4(k_1)-\frac{\mathbf{k}_2\cdot\mathbf{k}}{16}Y_4(k_2)+\frac{k^2}{24}\left(Y_4(k_1)+Y_4(k_2)\right)\right)+(D_1\leftrightarrow
D_2).
\end{align}
We are left with two
divergences. First, the one in $k^2 Y_4(k_i) $ which can be renormalized by a
multiplicative noise of the form $\tilde{J} \nabla^2 h$ and the one in $\mathbf{k}_i\cdot \mathbf{k}Y_4(k_i)$ which as
required by GI is the same as in the Power Spectrum.


\subsection{Local in Time}
The power counting of the effective viscosity and noise terms discussed above, corresponding to that employed in the EFToLSS literature, implies that loops are renormalizable if the effective terms are non-local in time. We will now demonstrate that by including the effective viscosity and the noise in the propagators, loops can also be renormalized in a theory local in time. In that case the theory is similar to an ordinary viscous fluid with stochastic noise.

Let us in the following write the viscosity as $\nu(D)=c_v^2\tilde{\nu}(D)$, with
$\tilde{\nu}(D)=O(1)$. We chose the time dependence of the noise to be
$\Delta(D,D')=\Delta \tilde{\nu}^3(D) \delta_D(D-D')$. The linear propagator
and the linear power spectrum are then given by
\begin{align}\label{prop-local}
G_0^R(D,D';k)&=e^{-c_v^2 k^2\int_{D'}^D d\eta\:
\tilde{\nu}(\eta)}\theta(D-D')\\
F_0(D,D';k)&=e^{-c_v^2 k^2\left(\int_{0}^D d\eta\: \tilde{\nu}(\eta)+\int_{0}^{D'}
d\eta\: \tilde{\nu}(\eta)\right)}P_{\Phi_{\rm in}}(k)+e^{-c_v^2
k^2\left(\int_{0}^D d\eta\: \tilde{\nu}(\eta)+\int_{0}^{D'} d\eta\: \tilde{\nu}(\eta)\right)}\Delta\int_{0}^{\rm{min}(D,D')} d\eta\:e^{2c_v^2 k^2\int_{0}^{\eta} d\eta'\: \tilde{\nu}(\eta')} \tilde{\nu}^3(\eta).\notag
\end{align}
As already argued in \cite{Rigopoulos2015a} on large scales we recover the usual linear Power Spectrum, while on small scales the Power Spectrum is dominated by the noise.
\begin{align}
F_0(D,D';k)&=e^{-c_v^2 k^2\left(\int_{0}^D d\eta\: \tilde{\nu}(\eta)+\int_{0}^{D'} d\eta\: \tilde{\nu}(\eta)\right)}\Delta\int_{0}^{\rm{min}(D,D')} d\eta\:\tilde{\nu}(\eta)e^{2c_v^2 k^2\int_{0}^{\eta} d\eta'\: \tilde{\nu}(\eta')} \tilde{\nu}^2(\eta)\notag\\
&=e^{-c_v^2 k^2\left(\int_{0}^D d\eta\: \tilde{\nu}(\eta)+\int_{0}^{D'} d\eta\: \tilde{\nu}(\eta)\right)}\frac{\Delta}{2c_v^2k^2}\int_{0}^{\rm{min}(D,D')} d\eta\: \tilde{\nu}^2(\eta)\frac{d}{d \eta}e^{2c_v^2 k^2\int_{0}^{\eta} d\eta'\: \tilde{\nu}(\eta')}\\
&=\frac{\tilde{\nu}^2(D')\Delta}{2c_v^2k^2}e^{-c_v^2 k^2\int_{D'}^D d\eta\: \tilde{\nu}(\eta)}\theta(D-D')+(D \leftrightarrow D') +O\left((c_vk)^{-4}\right).
\end{align}
To obtain the last line we performed a partial integration and neglected terms
which are exponentially suppressed. Repeated partial integration would allow to
calculate the $O\left((c_vk)^{-4}\right)$ terms and higher. Note that on small
scales the Power Spectrum is exponentially suppressed unless $D\approx D '$.
\footnote{More precisely it is suppressed unless $c_v^2 k^2\int_{D'}^D d\eta\:
\tilde{\nu}(\eta) \ll 1$, which translates for monotonic (growing)
$\tilde{\nu}(D)$ into $(D-D')<<\frac{1}{c_v^2k^2 \tilde{\nu}(D')}\sim
\frac{1}{c_v^2k^2}$.}

The leading UV-divergences of the self energies are then
\begin{align}
\Sigma\left(D,D';k\right)&\sim\frac{\Delta}{2c^2} \int \frac{d^3
q}{(2\pi)^3}
\frac{(q^2+\mathbf{q}\cdot\mathbf{k})\mathbf{q}\cdot\mathbf{k}}{q^2}\tilde{\nu}^2(D')\theta(D-D')
G_0^R(D,D';q)G_0^R(D,D';|\mathbf{q}+\mathbf{k}|)\notag\\
\Phi\left(D,D';k\right)&\sim\frac{\Delta^2}{4c^4} \int \frac{d^3
q}{(2\pi)^3}
\frac{(q^2-\frac{k^2}{4})^2}{(\mathbf{q}+\frac{\mathbf{k}}{2})^2(\mathbf{q}-\frac{\mathbf{k}}{2})^2}\tilde{\nu}^4(D')\theta(D-D')
G_0^R(D,D';|\mathbf{q}-\frac{\mathbf{k}}{2}|)G_0^R(D,D';|\mathbf{q}+\frac{\mathbf{k}}{2}|)+(D\leftrightarrow
D').
\end{align}
The above expressions are non-vanishing only for $D_1\approx D_2$ and they can therefore be approximated as local in time. Integrating over time we obtain to leading order
\begin{align}
\Sigma\left(D,D';k\right)&\sim\tilde{\nu}(D_2)\frac{\Delta}{2c^4}
\delta_D(D-D')\int \frac{d^3 q}{(2\pi)^3}
\frac{(q^2+\mathbf{q}\cdot\mathbf{k})\mathbf{q}\cdot\mathbf{k}}{q^2(q^2+(\mathbf{q}+\mathbf{k})^2)}\notag\\
\Phi\left(D,D';k\right)&\sim\tilde{\nu}^3(D_2)\frac{\Delta^2}{8c^6}\delta_D(D-D')
\int \frac{d^3 q}{(2\pi)^3} \frac{(q^2-\frac{k^2}{4})^2}{(\mathbf{q}+\frac{\mathbf{k}}{2})^2(\mathbf{q}-\frac{\mathbf{k}}{2})^2(q^2+\frac{k^2}{4})}.
\end{align}
We immediately see that the time dependence matches the one of the viscosity and
the noise.\footnote{Reference \cite{Rigopoulos2015a} treated the special case of
$\tilde{\nu}(D)=D$.} The $q$-integrals are actually
UV-divergent and must be regularized by an appropriate prescription. From
inspecting the integrals we see that $\Sigma$ has a quadratic and
a linear divergence. The quadratic vanishes due to rotational invariance and the
linear divergence has the correct $k$ dependence to be absorbed into the
viscosity. There is also a logarithmic divergence stemming form  the
sub leading terms $O(\frac{1}{q^4})$, we dropped this term since the divergence
vanishes. The linear divergence in $\Phi$ can be absorbed into
the noise.

Observe that the divergences are the same as in the time independent theory with $\tilde{\nu}=1$, as are the leading divergences at higher orders. There will also be additional divergences with new time dependencies at higher orders which will require new counter terms, including terms with additional time derivatives. This is not a problem since the theory is non-renormalizable and new counter terms
must be added any way. It is interesting to note that if instead of a scale independent noise we chose one with $\Delta\sim
k^{-1}$ in the UV, the theory becomes renormalizable by power counting.

If we follow the same procedure as for the self-energies for the triangle
diagrams we find
\begin{align}
\Pi(D,D_1,D_2;\mathbf{k}_1,\mathbf{k}_2)\sim -\frac{\Delta}{16
c^6}\delta_D(D-D_1)\delta_D(D-D_2
)\Bigg(&\int \frac{d^3
q}{(2\pi)^3}\frac{\mathbf{k}_1\cdot
\mathbf{q}\mathbf{k}_2\cdot(\mathbf{k}_1+\mathbf{q})\mathbf{q}\cdot(\mathbf{k}+\mathbf{q})}{q^2(q^2+(\mathbf{k}_1+\mathbf{q})^2)(q^2+(\mathbf{k}-\mathbf{q})^2)}\notag\\
+&\int
\frac{d^3 q}{(2\pi)^3}\frac{\mathbf{k}_1\cdot
\mathbf{q}\mathbf{k}_2\cdot
\mathbf{q}(\mathbf{k}_1+\mathbf{q})(\mathbf{k}_2-\mathbf{q})}{q^2(q^2+(\mathbf{k}_1+\mathbf{q})^2)(q^2+(\mathbf{k}_2-\mathbf{q})^2)}\Bigg)+\left(D_1\leftrightarrow
D_2\right)\notag\\
 \Psi(D,D_1,D_2;\mathbf{k}_1,\mathbf{k}_2)\sim
\frac{\Delta^2\tilde{\nu}^2(D)}{16c^6}\delta_D(D-D_1)\delta_D(D-D_2)
\Bigg(&\int
\frac{d^3q}{(2\pi)^3}\frac{\left(k_1^2-4q^2\right)\mathbf{k}\cdot\left(\mathbf{k}_1+2\mathbf{q}\right)\left(\mathbf{k}_1-2\mathbf{q}\right)\cdot\left(\mathbf{k}_1+2\mathbf{k}-2\mathbf{q}\right)}{\left(\mathbf{k}_1+2\mathbf{q}\right)^2\left(\mathbf{k}_1-2\mathbf{q}\right)^2}\notag\\
&\Bigg(\frac{1}{(k^2+4q^2)(k^2+4q^2+\mathbf{q}\cdot(\mathbf{k_1}-2\mathbf{q}))}\\
+&\frac{1}{(2k^2+(\mathbf{k}-2\mathbf{q})^2+\mathbf{k}\cdot(\mathbf{k}_1)-2\mathbf{q})((k^2+4q^2)+\mathbf{q}\cdot(\mathbf{k_1}-2\mathbf{q})}\notag\\
 +&\frac{1}{(k^2+4q^2)(2k^2+(\mathbf{k}-2\mathbf{q})^2+\mathbf{k}\cdot(\mathbf{k}_1)-2\mathbf{q})}\Bigg)\Bigg)+\left(D_1\leftrightarrow
 D_2\right)\notag,
\end{align}
The individual diagrams contributing to $\Pi$ are linearly divergent, but the
divergences cancel among the integrals, ensuring that the vertex is not
renormalized, as required by GI. $\Psi$ is by power counting logarithmically
divergent, but this divergence vanishes due to rotational invariance.

The third triangle diagram with three external dashed lines is finite and no
non-gaussian noise is required to obtain finite results. Similarly, higher order correlators are finite at one-loop, so the SAM and most likely also the fluid equations are one-loop renormalizable. Of course, the loop integrals still contain unphysical contributions from modes beyond a UV-cutoff $\Lambda\gg k_i$ and must in principle be renormalized, but the error made by not renormalizing the triangle diagrams is suppressed by $k_i/\Lambda$ and therefore of the same order as the residual cutoff dependence of renormalized self-energies, and hence only subleading. So the minimal set of counter terms required to make the local-in-time theory cutoff independent at leading order, and hence the number of free parameters, is smaller than in the non-local-in-time version. Note that finite contribution from the $g$- and $\lambda$-type vertices or new noise terms as discussed in section \ref{sec:Gal} can still be important but a detailed analysis is beyond the scope of our paper and we leave it for future work.


\subsection{Ward Identities}
The Ward Identities (WI) encode the statement that the effective action $\Gamma$
transforms under infinitesimal GT in the same way as the bare action $S$. This implies a set of relations that counterterms necessarily satisfy. The
effective action $\Gamma$ is related to the generating functional of connected
correlation functions and any physical information about a system can
be obtained from it. We will now briefly discuss how the WI ensure that in a
non-local theory the same divergences arise in different n-point functions.

The WI for the fluid equations were already derived in
 \cite{Peloso2013,Peloso2014}, see also \cite{Frey1994}. The corresponding WI
 for the SAM are
\begin{align}
\int dD d^3k\:  \Bigg(&
\mathbf{k}\cdot\mathbf{T}(D)\left(\frac{\delta\Gamma}{\delta h_{\mathbf{k}}(D)}h_{\mathbf{k}}(D)+\frac{\delta\Gamma}{\delta \chi_{\mathbf{k}}(D)}\chi_{\mathbf{k}}(D)\right)\notag\\
+&\delta_D(\mathbf{k})\left(\mathbf{\beta}(D)\cdot\partial_{\mathbf{k}}\frac{\delta\Gamma}{\delta
h_{\mathbf{k}}(D)}-\partial_D\mathbf{\beta}(D)\cdot\partial_{\mathbf{k}}\chi_{\mathbf{k}}(D)+\frac{i}{2}\chi_{\mathbf{k}}(D)\beta^2(D)\right)\Bigg)=0.\label{eq:WiAction}
\end{align}
The $\partial_{\mathbf{k}}\chi$-term relates the mean velocity to the Galilean boost
$\beta$ and the $\chi$-term is related to an unobservable shift in the potential. These terms correspond to the change of the bare action under GT, see equation \ref{eq:actionGT}. The 1-particle-irreducible vertices (1PI-vertices) defined as
\begin{align}
\Gamma^{(n,m)}&\equiv\frac{\delta^{n+m} \Gamma}{\delta
\chi_{\mathbf{p}_1}(\eta_1)\ldots\delta \chi_{\mathbf{p}_n}(\eta_n)\delta
h{\mathbf{q}_1}(\lambda_1)\ldots\delta h{\mathbf{q}_m}(\lambda_m)}\bigg
|_{h=\chi=0},\end{align}
are given by the sum over all 1PI-diagrams with $n$
dashed and $m$ solid lines.
Taking $n$ derivatives with respect to $\chi$ and $m$ derivatives with
respect to $\varphi$ of equation \eqref{eq:WiAction}, we arrive after a partial integration, for $n\neq1$
and $m\neq0$, at a relation between a vertex with
$n+m$ legs and one with $n+m+1$ legs of the form
\begin{align}
\sum_{i=1}^n \mathbf{p}_i \delta_D(D-\eta_i)
\Gamma^{(n,m)}+\sum_{i=1}^m\mathbf{q}_i \delta_D(D-\lambda_i)
\Gamma^{(n,m)}\notag\\
-\int d^3k\:
\delta_D(\mathbf{k})\partial_{D}\partial_{\mathbf{k}}\Gamma^{(n,m+1)}=0.\label{eq:wardindentity}
\end{align}
Note that $\Gamma^{(0,m)}=0$ for any $m$.
The WI in this form relate a 1PI-vertex to another 1PI-vertex with one more
leg.
The extra leg thereby corresponds to a velocity ie a solid line. The
delta-function picks out the limit $k\rightarrow 0$. Terms with more then one factor of $k$ are therefore unconstrained by the WI, as expected from our discussion in section \ref{sec:Gal}. Since each velocity leg must contain at least one factor of $k$ the WI relate 1PI-vertices with velocity legs with \emph{one} factor of $k$ to a 1PI-vertices with one velocity leg less. This is exactly what is required by GI as discussed in section \ref{sec:Gal}.

Let us illustrate that this ensures that loops can be renormalized with GI counter terms at any order in perturbation theory with the example of $2$-vertices and $3$-vertices. For $n=1$ and $m=1$ the WI read 
\begin{align}
\mathbf{k}_1 G^{-1}(D_1,D_2;k_1)\left(\theta(\eta-D_1)-\theta(\eta-D_2)\right)=\lim_{\mathbf{p}\rightarrow
0} \partial_\mathbf{k} \Gamma^{(1,2)}\left(D_1,D_2,\eta;\mathbf{k}_1,\mathbf{p}\right)\label{eq:WI11}\end{align}
where the full inverse propagator is given by
\begin{align}
G^{-1}(D_1,D_2;k_1)=\left(G_0^{-1}(D_1,D_2;k_1)-\Sigma(D_1,D_2,k_1)\right)
\end{align}
and similarly the full vertex is composed of the tree level vertex $\tilde{\gamma}$, containing both the usual vertex and a possible contribution from the non-local viscosity, and the loop corrections
\begin{align}
\Gamma^{(1,2)}\left(D_1,D_2,\eta;\mathbf{k}_1,\mathbf{k}_2,\mathbf{p}\right)=-2\tilde{\gamma}\left(D_1,D_2,\eta;\mathbf{k}_1,\mathbf{k}_2,\mathbf{p}\right)-\Pi\left(D_1,D_2,\eta;\mathbf{k}_1,\mathbf{k}_2,\mathbf{p}\right).
\end{align}
Using that in the local theory the tree level propagator is $G_0^{-1}(D_1,D_2;k_1)=\partial_{D_1}\delta_D(D_1-D_2)+k_1^2\nu(D_1)\delta_D(D_1-D_2)$ and the vertex is $\tilde{\gamma}=\delta(\eta-D_1)\delta(\eta-D_2)\frac{\mathbf{k}_1\cdot\mathbf{p}}{2}$ in equation \eqref{eq:WI11} we  find that convective derivative is GI. If the viscosity is non-local, its contribution does not cancel and the vertex gets an extra contribution, as discussed in section \ref{sec:Gal}.\\
For $n=2$ and $m=0$ a similar relation holds. At loop level we have 
\begin{align}
\Sigma\left(D_1,D_2;k_1\right)\mathbf{k}_1\left(\theta\left(\eta-D_1\right)-\theta\left(\eta-D_2\right)\right)&=\lim_{\mathbf{p}\rightarrow
0} \partial_\mathbf{p} \Pi\left(D_1,D_2,\eta;\mathbf{k}_1,\mathbf{p}\right)\notag\\
 \Phi\left(D_1,D_2;k_1\right)\mathbf{k}_1\left(\theta\left(\eta-D_1\right)-\theta\left(\eta-D_2\right)\right)&=\lim_{\mathbf{p}\rightarrow
 0}
 \partial_\mathbf{p}\Psi\left(D_1,D_2,\eta;\mathbf{k}_1,\mathbf{p}\right)\label{eq:WI2point}.
\end{align}
\begin{figure}
\centering
\includegraphics[width=0.99\textwidth]{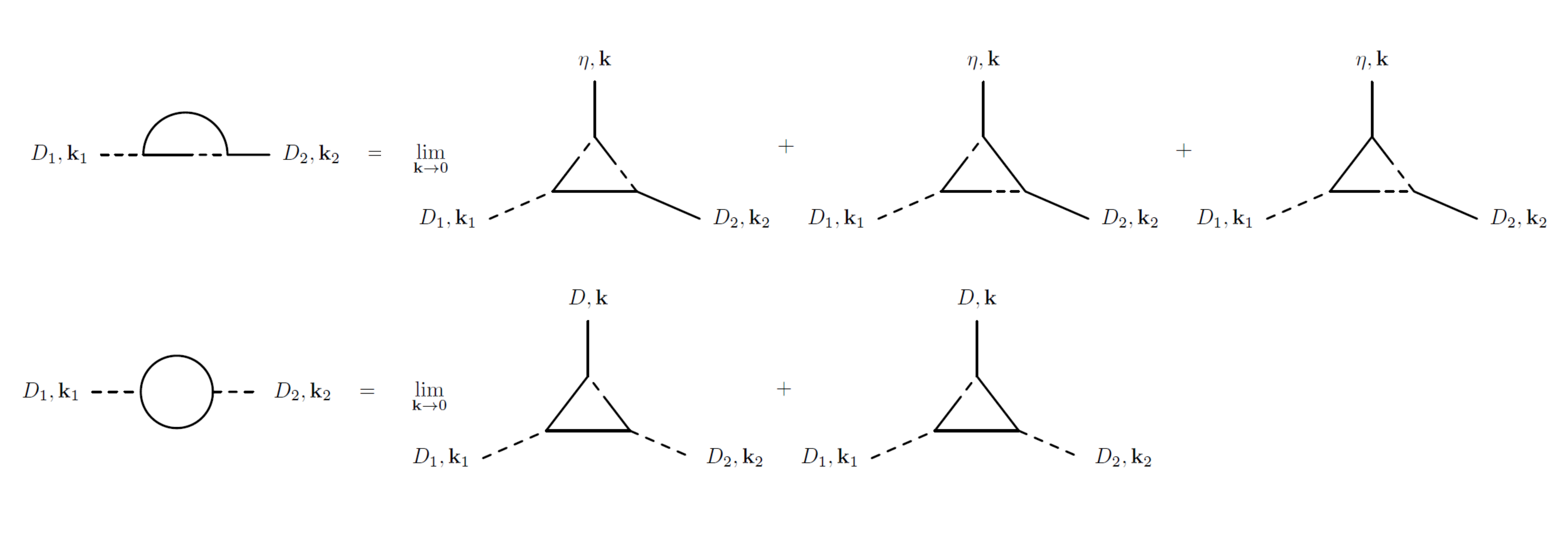}
\caption{The WI in equation \eqref{eq:WI2point} at one loop.}
\label{fig:WI}
\end{figure}
As an example the relations at one loop are depicted in fig. \ref{fig:WI}. If we
now use that in the limit $p\rightarrow 0$ the vertices are $\Pi(\mathbf{k}_1,\mathbf{p})\propto\mathbf{p}\cdot\mathbf{k}_1\Pi(k_1)$ and
$\Psi(\mathbf{k}_1,\mathbf{p})\propto\mathbf{p}\cdot\mathbf{k_1}\Psi(k_1)$, we
explicitly find the same relation as as we found in section \ref{sec:Gal}:
\begin{align}
\Pi(D_1,D_2,\eta;\mathbf{k}_1,\mathbf{p})&=\mathbf{p}\cdot\mathbf{k}_1\Sigma(D_1,D_2;k_1)\left(\theta(\eta-D_1)-\theta(\eta-D_2)\right)+O\left(p
^2\right)\notag\\
\Psi(D_1,D_2,\eta;\mathbf{k}_1,\mathbf{p})&=\mathbf{p}\cdot\mathbf{k}_1\Phi(D_1,D_2;k_1)\left(\theta(\eta-D_1)-\theta(\eta-D_2)\right)+O\left(p
^2\right).
\end{align}
These relations guarantee that the same divergence appearing in loop corrections
to the power spectrum also appears in loop corrections to the bispectrum. In
particular since the leading UV-divergences scale as $\Sigma\propto k^2$ and
$\Phi\propto k^0$, the WI guarantee that
the same divergence also appear in $\Pi$ and $\Psi$, if the theory is non-local
in time, while if the divergence is local in time the WI guarantee that there is
\emph{no} vertex correction of this type, unless the self energies contain time derivatives. Then as at tree level the WI guarantee that the time derivatives are always part of a convective derivative\footnote{These terms are redundant i.e. for practical calculations they can be simplified using the equation of motion.}.

From the form of the WI in equation \eqref{eq:wardindentity} it is immediately clear
that the same procedure also applies to higher order vertices. The general
relation between vertices with $n+m$ legs and $n+m+i$ legs can be obtained
by applying equation \eqref{eq:wardindentity} recursively.


\subsection{The Bispectrum in the local and non-local theories}
The scale and time dependence of correlation functions obtained from the local-in-time theory is much more complex then that obtained from the non-local-in-time one. But on
sufficiently large scales $c_v^2k^2\ll1$, and if the noises are neglected, the
one-loop Power Spectra of both theories are expected to agree. Nevertheless, the two theories give different results for the one-loop bispectrum.
In the non-local-in-time power counting, the one-loop bispectrum consists of the SPT tree-level and one loop bispectrum plus EFT
corrections in form of the new vertex, with the coupling $g$, and the counter terms for $\Sigma$:
\begin{align}
B_{\rm{NLT}}(D_1,D_2,D_3;k_1,k_2,k_3)&=B_{\rm{tree}}(k_1,k_2,k_3)+B_{\rm{1-loop}}(D_1,D_2,D_3;k_1,k_2,k_3)
+g
D_3^3\frac{\left(\mathbf{k}_1\cdot\mathbf{k}_2\right)^2}{6}P_{L}(k_1)P_{L}(k_2)\notag\\
&-c_v^2\frac{\mathbf{k}_1\cdot\mathbf{k}_2}{2}P_{L}(k_1)P_{L}(k_2)\left(\frac{k_1^2}{6}(D_3^3+3D_1^2D_3)+\frac{k_2^2}{6}(D_3^3+3D_2^2D_3)+\frac{k_3^2}{2}D_3^3\right)\\
&+\mathrm{permutations}\notag.
\end{align}
For the bispectrum in the local theory we find
\begin{align}
B_{\rm{LT}}(D_1,D_2,D_3;k_1,k_2,k_3)&=B_{\rm{NLT}}(D_1,D_2,D_3;k_1,k_2,k_3)-g
D_3^3\frac{\left(\mathbf{k}_1\cdot\mathbf{k}_2\right)^2}{6}P_{L}(k_1)P_{L}(k_2)\\
&+c_v^2D_3^3\frac{k_3^2\mathbf{k}_1\cdot\mathbf{k}_2}{12}P_{L}(k_1)P_{L}(k_2)
+\mathrm{permutations}\notag.
\end{align}
That $g$ appears in $B_{\rm{NLT}}$ but not in $B_{\rm{LT}}$ is an illustration that less free parameters are required for consistency in the local theory compared to the non local one. Of course, a $g$-type vertex can still be added to action but is not needed for renormalization. Similiar a $\lambda$-type vertex can be added to both the local and non-local theory, but is not required for renormalization.
The second term which is different comes from the extra
vertex enforced by GI and by integrals in the local theory of the form $\int_{0}^D d\eta\:\eta
f(\eta)$  which must be replaced by integrals of the form $\int_{0}^D
d\eta\:\int_{0}^{\eta'} d\eta'f(\eta')$ in the non-local theory. These integrals are
the same only if $f=1$, while for a power law $f\propto D^n$ both integrals become
$\propto D^{n+2}$ but with a different prefactor. Since for the full system the
time dependence of the propagator is more complicated one should expect that the
difference between $B_{\rm{LT}}$ and $B_{\rm{NLT}}$, the bispectra in the local and non-local theories, becomes larger within the
full theory.


\section{conclusion}
\label{sec:conclusion}
In this paper we discussed renormalization of the stochastic adhesion model (SAM). This is a toy model of structure formation, based on a simple parameterization of deviations from the Zel'dovich trajectories, which shares the same symmetries as the full set of Euler and continuity equations along with extra effective ``viscosity'' terms and a stochastic noise term. Because of these features we expect that the general conclusions arrived at here should also be applicable to the complete theory. 

Treating the viscosity terms as counter terms,  one is led through a one-loop calculation to a theory necessarily non-local in time. To ensure Galilean Invariance, non-local counter-terms must be evaluated along the fluid element path, introducing vertices at higher orders which are related to the lowest order counter terms by having the same coefficients. These terms are individually not GI, so their dependence on the wave-vector differs from that of local counter terms. As we have shown, these terms are required to ensure cutoff independence. This implies that the EFToLSS type approach with non-local-in-time counter terms is consistent with respect to renormalization, while the EFToLSS with local counter terms is not.

Alternatively, if the effective viscosity is included in the computation of the linear propagator \cite{Rigopoulos2015a, Blas2015}, leading to exponential damping on short scales, one obtains a consistent theory, renormalizable to one loop. For a numerically small viscosity one naturally recovers the known results with perturbative local counter terms for the one-loop Power Spectrum. In contrast to the EFToLSS type of approach though, no new counter terms are required to renormalize the one-loop bispectrum in this case. 

We briefly discussed the possibility to decide whether a theory local or non-local in time would be appropriate as an effective description of CDM by considering the bispectrum. As we pointed out, the precise size of corrections stemming from virialized scales are crucial in that respect. The local-in-time approach requires a minimal set of counter terms for renormalization 
and as a consequence the one-loop bispectrum is fully predicted by the one-loop Power Spectrum, although extra parameters are allowed. The non-local-in-time approach requires one more parameter for the renormalized bispectrum. We expect that similar conclusions would hold in the full system of continuity + Euler equations. Renormalized bispectra have only been computed in the EFToLSS approach which uses the non-local-in-time power counting. It would therefore be interesting to contrast the predictions of the local and non-local in time approaches in the full theory against the bispectrum from N-Body simulations.      

 \section*{Acknowledgments}
 F.F. acknowledges support by the IMPRS-PTFS.

\vspace{-5mm}
\bibliography{newrefs}

\end{document}